\documentstyle[12pt,amssymbols]{article}
\def\be{\begin{equation}}
\def\ee{\end{equation}}
\def\ba{\begin{eqnarray}}
\def\ea{\end{eqnarray}}

\def\mon{S^3\stackrel{S^1}{\rightarrow}S^2}
\def\inst{S^7\stackrel{S^3}{\rightarrow}S^4}
\def\octo{S^{15}\stackrel{S^7}{\rightarrow}S^8}
\def\fun#1#2{\lower3.6pt\vbox{\baselineskip0pt\lineskip.9pt
\ialign{$\mathsurround=0pt#1\hfill##\hfil$\crcr#2\crcr\sim\crcr}}}

\def\id{{\bf 1}}
\def\e{i\sigma}
\def\sm{$\sigma$-model}
\def\re#1{\cite{#1}}

\def\math{\mathsurround 0pt}
\def\oversim#1#2{\lower.5pt\vbox{\baselineskip0pt \lineskip-.5pt
        \ialign{$\math#1\hfil##\hfil$\crcr#2\crcr{\scriptstyle\sim}\crcr}}}

\def\d{\partial}
\def\half{{\mathchoice{{\textstyle{1\over 2}}}{1\over 2}{1\over 2}{1 \over 2}}}
\def\unit#1{\ifinner \;
            \else \quad \fi
            {\rm #1}}

\def\dbw{\mathop{\d}\limits^{\leftrightarrow}}
\def\tr{{\rm tr}}

\def\Z{{\Bbb Z}}
\def\R{{\Bbb R}}
\def\C{{\Bbb C}}
\def\Q{{\Bbb Q}}
\def\H{{\Bbb H}}

\begin{document}
\begin{titlepage}
\null\vspace{-62pt}
\begin{flushright}
NSF--ITP--92--93\\
CMU--HEP92--16\\
VAND--TH--92--10\\
DAMTP--HEP--92--61\\
hep-th/9209088 \\
June, 1992 \end{flushright}
\vspace{0.2in}
\begin{center}
{{\large \bf Generalized Semilocal Theories and Higher Hopf Maps}}\\
\vspace{0.3in}
{Mark Hindmarsh,$^{(a, b)}$\ \ Richard
Holman,$^{(a,c)}$\ \ Thomas W.~Kephart,$^{(a, d)}$\\
{\em and}\\
Tanmay Vachaspati$^{(a,e)}$\\}
\vspace{0.3in}
{{\it $^{(a)}$Institute for Theoretical Physics,\\
University of California, Santa Barbara CA~93106}} \\
\vspace{.1in}
{{\it $^{(b)}$Department of Applied Mathematics and Theoretical Physics,\\
University of Cambridge, Cambridge CB3~9EW}} \\
\vspace{.1in}
{{\it$^{(c)}$Physics Department, Carnegie Mellon University,
Pittsburgh, PA~15213}} \\
\vspace{.1in}
{{\it $^{(d)}$Department of Physics and Astronomy,\\
Vanderbilt University, Nashville, TN~37235}} \\
\vspace{.1in}
{{\it $^{(e)}$Tufts Institute of Cosmology, Department of Physics
and Astronomy, }}\\
{{\it Tufts University, Medford, MA~02155}}\\
\end{center}
\vspace{.4in}
\baselineskip=18pt
\vfill

\begin{abstract}
In semilocal theories, the vacuum manifold is fibered in a non-trivial way
by the action of the gauge group.  Here we generalize
the original semilocal theory (which was based on the Hopf bundle $\mon$)
to realize the next Hopf bundle $\inst$, and its extensions
$S^{2n+1}\stackrel{S^3}\rightarrow \H P^n$.  The semilocal defects in this
class of theories are classified by $\pi_3(S^3)$, and are interpreted as
constrained instantons or generalized sphaleron configurations. We fail to find
a field theoretic realization of the final Hopf
bundle $S^{15}\stackrel{S^7}{\rightarrow} S^{8}$, but are able to construct
other semilocal spaces realizing Stiefel bundles over Grassmanian spaces.
\end{abstract}
\end{titlepage}

\baselineskip=24pt

\vspace{24pt}

One of the most lively facets of the cosmology/particle physics interface
has been the utilization of topological defects arising from spontanously
broken symmetries in field theoretical models for cosmological
purposes \re{kib76}. Inflation was initially devised as a way of getting
around the inevitable existence of magnetic monopoles in grand unified
theories \re{MonoProb}, while cosmic strings \re{String} or cosmic
texture \re{Tex} may be of vital importance in understanding how large scale
structure is formed. The observation of the effects of topological defects
would give us a low energy window into the high energy physics regime.

It is sometimes assumed that the types of finite energy defects
supported by a given field theory were classified solely by the various
homotopy groups of the manifold of degenerate vacuua ${M} = G\slash H$,
where $G$ is the original symmetry group and $H$ the unbroken subgroup
\re{kib76}. However, care must be exercised when only a subgroup of $G$
is gauged, for in this case it is only the coset space of gauged
symmetries $G_l/H_l$ that is relevant \re{col}.  Indeed, Vachaspati and
Ach\'{u}carro \re{vac-ach} (VA) constructed a simple model exhibiting
finite energy stringlike configurations despite the vacuum manifold
having a trivial first homotopy group.  However, these so-called {\em
semilocal strings} (which are essentially Nielsen-Olesen vortices
\cite{NO}) are stable only when the Higgs mass is smaller than that of
the gauge boson \re{hin92a,hin92b}, from which we learn that the
usual interpretation of the
homotopic classification of defects fails for semilocal theories.
Preskill \cite{pre92} has emphasised that the surprising feature of
semilocal strings is that they are unstable for sufficiently large
Higgs masses, and has also extended the analysis to models which
possess approximate global symmetries over and above the exact ones.
These would arise through gauge boson loops reducing an accidental
symmetry of the potential not possessed by the whole Lagrangian,
and can act to stabilize otherwise unstable defects.
These theories have other fascinating properties \cite{gib+,abr} and
may have important consequences for cosmology \cite{hin92b}.   Thus
they merit thorough investigation.

The original semilocal theory \cite{vac-ach,hin92a,hin92b} is one of
$(n+1)$ complex scalar fields with a global $U(n+1)$ symmetry, of which the
abelian $U(1)$ is gauged  (this was termed an extended Abelian Higgs
(EAH) model in references \cite{hin92a,hin92b}).  The unbroken global
symmetry is $U(n)$, and $H_l$ is trivial.  Thus the vacuum manifold
is $U(n+1)/U(n) \simeq S^{2n+1}$, which is fibred by the $S^1$ gauge
orbits in a non-trivial way:  the base space is $\C P^n$
\cite{hin92a,hin92b}.  At low momenta (lower than the masses of the
Higgs and vector particles) this base space is the target space of a
non-linear \sm, and the coordinates of this space are the
$2n$ Goldstone modes of the
theory.  The gauge coset space $M_l$ is $G_l/H_l \simeq S^1$, showing
that the theory has non-trivial $\pi_1$ and therefore the possibility
of vortex configurations.

In this paper we extend the class of semilocal theories in several
directions, but first we review the relation between the vacuum manifold
of the $n=1$ EAH model
 and the Hopf bundle $\mon$. This discussion
then sets the stage for the main work of this paper, which is the
extension of the original semilocal theory to a class of field theories
whose first member provides a model for the higher Hopf bundle $\inst$.
The low momentum dynamics of this theory will turn out to be that of a
{\em quaternionic} ${\H P}^n$ non-linear \sm\ \re{gal}.
We will then elucidate the nature of the
defect found in this theory.  There are a couple of other possible
generalizations.  One of these is to exploit the fact that $\C P^n
\simeq SU(n+1)/[SU(n)\times U(1)]$ is one member of the set of complex
Grassman manifolds, $\C G(n,m) \simeq SU(n+m)/S[U(n)\times U(m)]$, and
recast the \sm\ construction \cite{bre-hik-zin} into that of a spontaneously
broken gauge theory.  We can also do this for the real Grassman
manifolds $\R G(n,m) =
SO(n+m)/S[O(n)\times O(m)]$.  However, the step up from
quaternions to octonions does not seem to provide us with yet another class.

The semilocal string model of VA is based on the symmetry group $G =
[SU(2)_g\times U(1)_l]/\Z_2$, where the subscripts $g, \ l$ indicate whether
the symmetry is global or local, respectively. $G$ is then broken down
to $H = U(1)_g$ by a {complex} doublet $\Phi$. The generator of $H$ is
a linear combination of the diagonal generator of $SU(2)_g$ and the
$U(1)_l$ generator.  This is essentially the scalar sector of the
electroweak model in the limit that the $SU(2)_L$ gauge coupling $g$ is
set to zero, a fact which was used by Vachaspati to point out the
possibility of classically stable vortex solutions in the Standard
Model \cite{vac92}.

Let us now look for topological defects in this theory. This means that
we are looking for static solutions of the field equations which are
separated from the vacuum $|\Phi| = 0$ by field configurations whose
energy diverges proportionally to the volume in the infinite volume
limit.  These solutions need not have finite energy themselves, but
typically one requires that their energy diverges less fast than the
volume.  The energy functional in $D$ spatial dimensions for static
configurations is
\be
{\cal E} =  \int d^Dx \left[\frac{1}{4}F_{ij}F_{ij}
+ |D_i\Phi|^2 + V(\Phi)\right]
\ee
where the Higgs potential is $V(\Phi) = \half\lambda(\Phi^{\dag} \Phi -
\eta^2)^2$.   The requirement that the energy diverge less fast than
the volume $R^D$ means that $\Phi$ must approach the manifold of zeroes
of $V$, the vacuum manifold $M$, faster than $|x|^{-D}$ at infinity.
$M$ is given by $\Phi^{\dag} \Phi = \eta^2$, which is easily seen to be
$S^3$ when $\Phi$ is written in terms of the four real fields that are
the real and imaginary parts of each component. Now both $\pi_1(S^3)$
and $\pi_2(S^3)$ are trivial, so we might not expect to have either
cosmic strings or monopoles arising from this symmetry breaking.
However, this conclusion is premature.
Let us examine the covariant derivative contribution to the energy
functional.  Naively, this has a potential divergence going as
$R^{D-2}$.  However, if $\Phi$ is a gauge transformation at infinity,
this term can be finite.  Thus requiring finite energy of a field
configuration necessarily means that $\Phi$ must lie on a gauge orbit,
which has the topology of the coset space $M_l \subset M$.  Thus if
$\pi_{D-1}(M_l)$ is non-trivial, we should expect to find finite energy
defects.  Finally, the divergence of the contribution of the gauge
field to the total energy depends on how fast $F_{ij}$ tends to zero,
which in turn depends on how fast the currents producing the field
vanish.  This is a dynamical rather than a topological question.

In the case at hand, $U(1)_l \rightarrow 1$, so that the gauge part of
the vacuum manifold is isomorphic to $S^1$.  Every point on the $S^3$
defined by the constraint on the vacuum value of $\Phi$ lies on an
$S^1$ defined by the action of $U(1)_l$ on $S^3$ ({\em i.e.} the action
of $G_l$ is effective on $M$).  It costs no energy to move along one of
these circles in field space. In 2 space dimensions we can therefore
wrap the field at spatial infinity $\Phi_\infty$ around any of these
$S^1$'s, giving rise to an infinite number of distinct field
configurations. Extremizing in one of these sectors in field
configuration space should give rise to a string solution.
Now since $S^3$ is simply connected, it would be tempting to argue that
these strings are deformable to the vacuum ($\Phi_\infty$ constant)
simply by shrinking the loop defined by $\Phi_\infty$ to a point in
$M$, and thus the strings will be unstable. The catch in this argument
is that in order to perform this deformation $\Phi_\infty$ must leave
the set of gauge orbits, so that at infinity it has a component which
is a {\em global} $SU(2)_g$ transformation.  Thus we pay a
logarithmically divergent penalty in gradient energy. There is, then,
an infinite energy barrier separating the vortex configurations from
the vacuum.

The non-trivial embedding of the gauge vacuum manifold $S^1 =
U(1)\slash 1$ in the full vacuum manifold $S^3$ tells us that what we
are dealing with is a fiber bundle structure where the total space is
$S^3$, the fiber $S^1$ and the base space is the coset of $S^3$ by the
$S^1$ action, i.e. $S^2$. This is nothing but one of the celebrated
Hopf fibrations of spheres by spheres \cite{FibBun}, sometimes termed
the monopole bundle \cite{egu-gil-han}.
The fact that the space obtained by modding out by
the action of $S^1$ on $S^3$ is $S^2$ also indicates that the theory
supports global monopole configurations. These monopoles are at the end
of the strings \cite{hin92b,gib+}.

In the general EAH model the symmetry group is $[SU(n+1)_g\times
U(1)_l]/\Z_{n+1}$,  $n>1$, which is spontaneously broken by a complex
$(n+1)$-plet
to $U(n)_g\sim [SU(n)_g \times U(1)_g]/\Z_n$
 \re{hin92a,hin92b}. The vacuum manifold in this case is $S^{2n+1}$,
 while the gauge orbits are again just $S^{1}$. Thus, despite the fact
that for $n>1$, $\pi_1(S^{2n+1})$ is trivial, this theory will still
support semilocal strings. Furthermore, the low momentum limit of this
theory (when mean momenta are much less than the scalar and vector
masses) can be described by the dynamics of the ${\C P}^{n}$ nonlinear
$\sigma$-model. From a bundle point of view, the above construction
just defines the following extension of monopole Hopf bundle: $S^{2n+1}
\stackrel{S^1}{\rightarrow} {\C P}^{n}$.

The other way of constructing semilocal defects is to make use of the
other Hopf bundles $\inst$ and $\octo$ and find field theoretic models
that realize them as in the monopole bundle case. What we will find is
that only the first of these, often called the instanton bundle,
lends itself to such a realization.  This is
essentially because $S^3$ is also the group $SU(2)$, while $S^7$ is not
a group at all.

There are a variety of equivalent ways of approaching the construction
of the field theoretic model of the instanton Hopf bundle. For example,
it is well known \re{FibBun} that the various Hopf bundles are
intimately related to the various division algebras (reals, complex
numbers, quaternions and octonions), with the monopole and instanton
bundles being related to the complex numbers and quaternions,
respectively.  So let us try replacing the complex $(n+1)$-vector
$\Phi$ by $(n+1)$ quaternions $Q \equiv (q_1,...,q_{n+1})$.  The group
of symmetries that leave the norm $|\Phi|^2$ invariant is $U(n+1,\C)$.
Analagously, the quaternionic scalar product is invariant under the
group $U(n+1,\Q)$ \cite{gil}, which is the group of unitary rank $n+1$
matrices with quaternionic entries.  This group is isomorphic to the
symplectic group of rank $n+1$, $Sp(n+1)$ \cite{gil}.

We may represent the components of $Q$ by $2\times 2$ matrices, using
the Pauli matrices as a basis for the nonreal elements:
\be
q_a = q_a^0\id + q_a^A(\e^A)
\ee
with $q_a^0$ and $q_a^A$ ($A\in\{1,2,3\}$) real.  The scalar product is
then defined with the usual matrix trace:
\be
(Q,Q) \equiv \half\sum_1^{n+1}\tr(q_a^{\dag} q_a) =
\sum_1^{n+1}[(q_a^0)^2 + (q_a^i)^2]
\ee
The action of an element $M_{ab}$ of $U(n+1,Q)$ can be chosen to be by
right multiplication, so that
\be
q_a \to q'_a = q_bM_{ab}
\ee
But there is also another symmetry of the scalar product, which is $q_a
\to q'_a = sq_a$, with $s^{\dag} s = \id$.  Because we are representing
the quaternions by matrices, it is clear that this group of symmetries
is $SU(2)$, or equivalently $Sp(1)$.  The full symmetry group is not
quite the direct product of the two groups, because $s=-\id$ has the
same action as $M_{ab} = -\id\delta_{ab}$.  Therefore, for this
quaternionic extension we have
\be
G = [Sp(n+1)_g\times Sp(1)_l]/\Z_2
\ee
In the original
semilocal theory, the overall complex phase of the scalar field was
gauged.  Thus the quaternionic analogy is to gauge an overall
quaternionic phase -- the $Sp(1)$ subgroup which acts from the left.
The covariant derivative is therefore
\be
D_\mu Q = (\d_\mu - gA_\mu)Q \qquad (A_\mu = \half\e^A A_\mu^A)
\ee
We can now write down the Lagrangian of the model:
\be
{\cal L} = -\half (F_{\mu\nu},F^{\mu\nu}) + (D_\mu Q,D^\mu Q) -
\half\lambda\left((Q,Q) - \eta^2\right)^2
\ee
where $F_{\mu\nu} = \d_{(\mu}A_{\nu)} - q[A_\mu,A_\nu]$.  The potential
is stable to radiative corrections, since the whole Lagrangian has been
designed to be invariant under the symmetry group $G$.

The vacuum manifold $M$ of this theory is $S^{4n+3}$, because the
potential constrains the sums of the squares of $4(n+1)$ real numbers to be
constant.  The gauge orbits are 3-spheres, for they preserve the
magnitude of $(q_a^0)^2 + (q_a^i)^2$ for each $a$.  The unbroken
subgroup consists entirely of global symmetries and is $[Sp(n)\times
Sp(1)]/\Z_2$.  This can be ascertained by choosing the vacuum
expectation value to be $Q_0 = (\eta\id,0,...,0)^T$, which is always
possible since we have the freedom to make global $U(n+1,\Q)$ rotations
without changing any of the physics.  The unbroken $Sp(n)$ subgroup
acts on the lower $n$ components of $Q_0$, while the unbroken global
$Sp(1)$ is the diagonal subgroup of $Sp(n+1)\times Sp(1)$ consisting of
elements of the form
\be
h = s^{-1}\delta_{ab}\otimes s
\ee
This is the quaternionic analogue of the unbroken global $U(1)$ in the
complex field case.

We now demonstrate that the low momentum dynamics of this theory are
those of a \sm\ with target space $\H P^n$.  First, we recall the
definition of this class of quaternionic projective spaces
\cite{gil}.  The manifold $\H P^n$ is constructed from
$\H^{n+1}-\{0\}$ (the set of non-zero quaternionic $(n+1)$-vectors
$Q$) by identifying all elements which are equivalent by left
multiplication:  {\em i.e.}, $Q\equiv Q'$ if and only if there is a
quaternion $p$ such that $Q' = pQ$.  This exactly parallels the
construction of $\C P^n$.  Equally, we can take as a representative class
the set of all unit vectors, in which case we identify all elements of
that set which are related by left multiplication by unit quaternions.
Recalling that the unit quaternions form a group isomorphic to $SU(2)$,
we may write $S^{4n+3} \stackrel{S^3}{\rightarrow} \H P^n$.

For low momenta, the gauge field is entirely determined by the
gradients of the scalar field:
\be
A_\mu^A = -{1\over 2g} {(Q\e^A, \dbw_\mu Q) \over (Q,Q)}
\ee
while the scalar field keeps to its vacuum manifold.  Thus the
effective lagrangian at low momentum is
\be
{\cal L}_{HP} = \half\sum_a\left[\tr(\d_\mu q_a^{\dag}\d^\mu q_a) -
\tr(\d_\mu q_a^{\dag} q_a)\tr(q_a^{\dag}\d^\mu q_a)\slash
\sum_a\tr(q_a^{\dag} q_a)\right]
\ee
which is that of the $\H P^n$ \sm\ with the Fubini-Study metric
\cite{gal,gur-tze}.

The use of quaternions may seem like unnecessary obfuscation.  However,
there is a representation of this theory in terms of $n+1$ complex
doublets $\phi_a$, since each quaternion may be written
\be
q_a \equiv (\phi_a \ i\sigma^2\phi_a^*)
\ee
by which we mean that each entry in the brackets is a column in the
matrix representing the quaternion.  Thus we are in essence considering
a model which $n+1$ copies of the scalar sector of the Electroweak
theory, equipped with a $Sp(n+1)$ global symmetry,  in the limit
$\theta_W=0$.

A case of particular interest is the smallest member of the new class,
where $n=1$, and so we will examine it in detail.  Here we have
the symmetry group $Sp(2)_g\times Sp(1)_l \simeq Spin(5)_g\times
SU(2)_l$ \re{gil}.  The representation of the global group in which
the scalar field lies is the fundamental of $Sp(2)$ i.e. a (pseudo)
real ${\bf 4}$, and that of the local group is the fundamental of
$Sp(1)$. Thus $\Phi \equiv (\phi_1,\phi_2)^T$ transforms as a $({\bf
4}, {\bf 2})$ of $Sp(2)_g\times Sp(1)_l$.

We now analyze how the representations decompose in the symmetry
breaking.  First, note that $Sp(1)\times Sp(1)\subset Sp(2)$ (this is
just the $Spin(4)\subset Spin(5)$ subgroup chain). Under this subgroup,
the ${\bf 4}$ of $Sp(2)$ decomposes as ${\bf 4} \rightarrow ({\bf 1},
{\bf 2}) \oplus  ({\bf 2}, {\bf 1})$. Thus, when one of these
components acquires a vacuum expectation value, one of the $SU(2)$'s
will be broken while the other remains untouched. However, at the same
time, the {\em local} $SU(2)_l$ is also broken via a doublet. Thus just
as in chiral $SU(2)\times SU(2)$ models, the generators of both of the
broken $SU(2)$'s can be combined so as to form an unbroken {\em global}
$SU(2)_g'$. The symmetry breaking pattern is then just $Sp(2)_g\times
SU(2)_l\rightarrow Sp(1)_g\times SU(2)_g'$. We see from this that the
vacuum manifold is of dimension $(10 + 3) -(3 + 3) = 7$. Furthermore,
one can think of $Sp(2)\simeq Spin(5)$ as the product $S^{7}\times S^{3}$,
while the unbroken subgroup is just $S^3 \times S^3$. Thus, by
``cancelling'' the $S^3$'s in forming the coset, we see that the vacuum
manifold should be $S^7$. This is, of course, obvious from the Higgs
potential. The only renormalizable $Sp(2)_g\times SU(2)_l$ invariant
potential is
$
V(\Phi) = \half\lambda (\phi_1^{\dag} \phi_1 + \phi_2^{\dag} \phi_2 - \eta^2)^2
$,
which contains eight real fields.  It is clear from the form of
$V(\Phi)$ that the vacuum manifold is $S^7$.  However, just as in the
$n=1$ EAH model, the broken gauge symmetries $SU(2)_l \simeq S^3$ act
non-trivially on this $S^7$, yielding the Hopf bundle $\inst$.

Thus the low momentum effective \sm\ has a target space $S^4$, and it
is no surprise that the Lagrangian can be rewritten in terms of an
$O(5)/O(4)$ \sm . This parallels the equivalence between $\C P^1$ and
$S^2$ \cite{raj,hin92b}.  We can make use of the isomorphism between
$Sp(2)$ and $Spin(5)$, and utilize the 5 gamma matrices
$\Gamma^m$ to construct 5 real fields $\varphi^m$ as follows:
\be
\varphi^m = (Q,\Gamma^mQ)/\eta
\ee
Recalling the relation
\be
\sum_m \Gamma^m_{\mu\nu}\Gamma^m_{\rho\sigma} = {4\over 3} \delta_{\mu\sigma}
\delta_{\nu\rho} - {1\over 3} \delta_{\mu\nu}\delta_{\rho\sigma}
\ee
we find that in terms of the 5 real fields the \sm\ Lagrangian is
\be
{\cal L}_{O(5)} = {3\over 8}\sum_m \d_\mu\varphi^m\d^\mu\varphi^m
\ee
(supplemented by the constraint $\sum_m(\varphi^m)^2 = \eta^2$).

Let us now consider the finite energy, or more appropriately, finite
Euclidean action, topological defects in this class of models.  In $D$
dimensions the action functional is
\be
S = \int d^Dx \left[\half(F_{\mu\nu},F_{\mu\nu}) +
(D_\mu Q, D_\mu Q) + V(Q) \right]
\ee
where $\mu$ runs from 1 to $D$.  Each term on the right hand side must
vanish separately faster than $|x|^{-D}$ for a finite action solution,
which means that not only must $Q$ lie in $M$ at infinity, but also
that it be a gauge transform of the vacuum:
\be
Q = \Omega(\hat{x}_\mu) Q_0
\ee
where $\Omega(\hat{x}_\mu)$ is an $Sp(1)$ matrix.  Thus at infinity,
$Q$ must lie entirely in one of the gauge orbits, which are 3-spheres.
The finite action configurations on this theory are therefore
classified by $\pi_{D-1}(S^3)$, not by the homotopy groups of the
vacuum manifold, $S^{4n+3}$.  Of particular interest is the 4
dimensional case, for which the finite action field configurations fall into
homotopically inequivalent classes labelled by $\pi_3(S^3) = \Z$.  The
non-trivial elements correspond to instantons \cite{raj}.  The instanton {\em
solutions} to the field equations following from this action exist only
for $\eta=0$, when the conformal symmetry of the pure Yang-Mills
theory is re-established \cite{tho}.
Nonetheless, instanton {\em configurations}
are important in the quantum theory, where the path integral ensures
that classically disallowed configurations contribute to physical
quantities.  In the $SU(2)_l$ theory with one scalar doublet, the
instantons that contribute the most are those that are stable to all
perturbations except dilatations, which near the origin have the form
\cite{aff}
\be
q_1 = {e_\mu x_\mu \over (x^2 + \rho^2)^\half} \qquad gA_\mu^A =
{\eta^A_{\mu\nu}x_\nu \over (x^2 + \rho^2)^\half}
\ee
Here, $\eta^A_{\mu\nu} = -i \tr(e_\mu^{\dag}\sigma^Ae_\nu)$
\cite{tho}.  Note
that the Higgs field has to vanish at the centre of the instanton,
because on the 3-sphere at infinity ($S^3_\infty$) it wraps exactly once around
its vacuum manifold,  which is topologically  $S^3$.  However, in our
semilocal extension to this $SU(2)$ model, there is no need for the
other $SU(2)$ doublets to vanish at the origin.  Indeed, for very large
instantons we know that the theory is approximately an $\H P^n$ \sm , and
thus the field will stick close to its target space everywhere.  Since
the scalar field must be a gauge transformation at infinity, it maps
$S^3_\infty$ to a single point in $\H P^n$, while over the rest of the
spacetime $\R^4$ it completes a non-contractible 4-sphere in $\H P^n$.
The case $n=1$
is simple to deal with, for there $\H P^1 \simeq S^4$, and as a
result we have the luxury of being able to construct the instanton
configurations in an obvious way out of the real fields $\varphi^m$.
Large instantons constrained to have scale size $\rho$ will take the
form
\be
\varphi^\mu = \sin\psi(x/\rho)\hat{x}_\mu \qquad \varphi^5 = \cos\psi(x/\rho)
\ee
with $\psi(0)=0$ and $\psi(\pi)=\pi$.  The exact form of $\psi(x/\rho)$
will depend on how the instantons are constrained \cite{aff}.

Similar remarks apply to the sphalerons \cite{Sphal} in the model.  In the
construction of the sphaleron of the 1-doublet theory, the 2-sphere
at spatial infinity covers the entire vacuum manifold as the loop
parameter $\tau$ varies between 0 and 1.  The continuity of the
field then implies that somewhere in $\R^3\times [0,1]$ the field must
vanish.  When there are $n$ other scalar doublets, there is no
topological argument forcing the other fields to vanish if their
$SU(2)_l$ orientation remains fixed.  Whether or not the scalar fields
actually do vanish at the centre of the sphaleron is a dynamical
question, as for the semilocal string.

To summarise the results so far:  we have constructed another class of
semilocal theories realising the fibrations $S^{4n+3}
\stackrel{S^3}{\rightarrow} \H P^n$, the first member of this class being
the higher Hopf map $\inst$.  Emboldened by this success, one might be
tempted to try and realise the highest Hopf map $\octo$, and its
companion octonionic projective space $F_4/Spin(9)$.
However, we
have been unable to do this in the current framework of a spontaneously
broken theory with gauge and global symmetries, for the
canonical semilocal construction
fails for this Hopf fibration.  The natural thing to do is to
again duplicate the scalar field content.  Thus we would have 4
quaternionic fields, or equivalently 2 fields $Q_1$ and $Q_2$ in the
{\bf 4} of $Sp(2)$.  If we gauge this $Sp(2)$, then the breaking $Sp(2)
\to Sp(1)$ gives us the required gauged $S^7$.  In order to make the
full vacuum manifold $S^{15}$, we must use the $O(16)$  symmetric
potential $\lambda[(Q_1,Q_1)+(Q_2,Q_2) - \eta^2]^2$.  However, this
does not fix the relative orientation of $Q_1$ and $Q_2$, which is
necessary to ensure an unbroken gauged $Sp(1)$.  This would entail
introducing a term $\lambda'(Q_1,Q_2)(Q_2,Q_1)$, which reduces the dimension
of the vacuum manifold.  We have been unable to prove that a semilocal
theory realising this fibration does not exist, but we have not
succeeded in finding one.

However, we can generalise the existing semilocal theories in other
directions.  For example, $\C P^n$ is one member of the family of complex
Grassmanian manifolds $\C G(n,m) = SU(n+m)/S[U(n)\times U(m)]$ \cite{gil}.
This
suggests constructing a theory in which a $U(m)$ gauge symmetry is
completely broken by $n+m$ scalars in the fundamental of $U(m)$.  Let
us denote them by $\phi_{a}$, where $a=1,\ldots,n+m$.  Drawing on
the Grassmanian \sm\ construction of Br\'ezin et al \cite{bre-hik-zin},
it is easy to
construct the correct potential that gives us the required
symmetry breaking:
\be
V = \half\lambda(\bar{\phi}^a\phi_a - m\eta^2)^2
 + \half \lambda'(\bar{\phi}^a\phi_b - \eta^2\delta^a_b)^2
\ee
When $\lambda'>0$ this ensures that the vacuum solution satisfies
\be
\bar{\phi}^a\phi_b = \eta^2P^a_b
\ee
where $P^a_b$ is a rank $m$ projector onto the broken part of the
global $SU(n+m)$ symmetry.  We can always choose a basis in which
$\phi_{ia}$ = $\delta_{ia}$ for $i,a = 1,\ldots,m$, so that it is clear
that the local $U(m)$ mixes with a $U(m)$ subgroup of the global
symmetries to form an unbroken global subgroup.  Thus the symmetry
breaking pattern is
\be
SU(n+m)_g \times U(m)_l \rightarrow SU(n)_g \times U(m)_g
\label{e:break}
\ee
The full vacuum manifold $M$ is the space of $m$-frames in $\C^{n+m}$,
otherwise known as the Stiefel manifold $V_{n+m,m}(\C)$, which is
isomorphic to $SU(n+m)/SU(n)$.  This space is fibred by the $U(m)$
gauge orbits, which must be factored out to obtain the low energy
\sm\ target space $M_g$.  Thus we realize the bundle
\be
V_{n+m,m}(\C) \stackrel{U(m)}{\rightarrow} \C G(n,m).
\ee

Grassman manifolds exist over the other commutative division algebras
$\R$ and $\Q$ \cite{gil}, and they can be constructed in a semilocal
context by replacing the complex fields of the preceding discussion
by real or quaternionic ones respectively.  This has the effect of
changing the unitary symmetries into orthogonal or symplectic ones,
and the symmetry breaking
patterns become
$$
\displaylines{
SO(n+m)_g \times SO(m)_l \rightarrow SO(n)_g \times SO(m)_g\cr
Sp(n+m)_g \times Sp(m)_l \rightarrow Sp(n)_g \times Sp(m)_g\cr
}
$$
Preskill \cite{pre92} has constructed an example of a semilocal theory
which in a sense mixes the real and complex cases.  One can choose
to gauge only an $SO(m)$ subgroup of the $U(m)$ symmetry in the
$\C G(n,m)$ model, producing the breaking
\be
U(n+m)_g\times SO(m)_l \rightarrow U(n)_g\times SO(m)_g
\ee
When $n=0$ and $m=3$ we obtain Preskill's model.  All
the models over $\R$ and $\C$
have topologically non-trivial vortex configurations,
since $\pi_1(G_l)$ is non-trivial for $G_l \simeq SO(m)$ and  $U(m)$.
When the gauge group is $SO(2)$ or $U(m)$, the vortices are labelled by
an integer, because of the broken $U(1)$ factor.
When the gauge group is $SO(m)$
($m>2$), the first homotopy group is $\Z_2$.

Lastly, we point out that there are quaternionic projective spaces
which are not Grassmanian.  They are finite in number, and involve the
exceptional groups \cite{gal}:
$$
\displaylines{ G_2/SU(2)\times Sp(1) \qquad F_4/Sp(3)\times Sp(1)
\qquad E_6/SU(6)\times Sp(1)\cr
E_7/Spin(12)\times Sp(1)\qquad E_8/E_7\times Sp(1)\cr}
$$
It would be interesting to construct semilocal models which realise
these manifolds as their \sm\ target spaces.

To conclude, we have constructed several more classes of semilocal
model, all of whose low energy \sm\ target manifolds are
Grassman manifolds.
Any GUT with continuous global
symmetries must confront the issue of semilocality, so these
investigations are expected to have some importance. Their value lies
particularly in the sphere of the cosmological predictions of the
theory: for example, a semilocal theory may not have the stable cosmic
strings that the gauge sector symmetry breaking may predict; or the
strings may terminate on global monopoles \cite{hin92b,gib+}.
We hope to return to the
subject of semilocal GUTs in a future publication.

 This research was supported in part by the National Science Foundation
 under grant No. PHY89--04035. RH was supported in part by DOE contract
DE--FG02--91ER40682. TWK was supported by the DOE (grant
DE--FG05--85ER40226).  MH wishes to thank the Theoretical Division at
the Los Alamos National Laboratory and the Theoretical Astrophysics
group at Fermilab for their hospitality while this work was being completed.


\vspace{36pt}
\begin{thebibliography}{99}

\def\ref #1 #2 #3 #4 #5 #6{#1, {\it #2} {\bf #3}, #4 (#5)#6\ }
\def\pl{{Phys. Lett.}}
\def\prl{{Phys. Rev. Lett.}}
\def\pr{{Phys. Rev.}}
\def\cqg{{Class. Quantum Grav.}}
\def\apj{{Ap. J.}}
\def\np{{Nucl. Phys.}}
\def\cmp{{Commun. math. Phys.}}
\def\jp{{J. Phys.}}
\def\rmp{{Rev. Mod. Phys.}}
\def\prep{{Phys. Rep.}}
\def\rpp{{Rep. Prog. Phys.}}
\def\jmp{{J. Math. Phys.}}

\bibitem{kib76} \ref{T. W. B. Kibble} {\jp} {A9} 1387 1976 .
\bibitem{MonoProb} \ref{Ya. B. Zel'dovich and M. V. Khlopov}
{\pl} 79B 239 1978 ; \ref{J. Preskill} {\prl} 43 1365 (1979) .
\bibitem{String} \ref{A. Vilenkin} {\prep} 121 1 1985 ;
N. Turok in ``Particles, Strings and
Supersymmetry'' A. Jevicki and C-I. Tan Eds. (World Scientific, Singapore,
1989); \ref{R. Brandenberger} {\jp} G15 1 1989 .
\bibitem{Tex} \ref{N. Turok} {\prl} 63 2625 1989 ; \ref{
D. Spergel and N.Turok} {\prl} 64 2736 1990 ;
\ref{D. Spergel, N. Turok, W. Press and B. Ryden} {\pr} D43 1038 1991 .
\bibitem{col} S. Coleman, ``Aspects of Symmetry'' (CUP, Cambridge, 1985).
\bibitem{vac-ach} \ref{T. Vachaspati and A. Ach\'ucarro}
{\pr} D44 3067 1991 .
\bibitem{NO} \ref{A. A. Abrikosov} {Sov. Phys. JETP} 5 1174 1957 \relax
[{\it Zh. Eksp. Teor. Phys.} {\bf 47}, 2222 (1957)];
\ref{H. Nielsen and P. Olesen} {\np} B61 45 1973 .
\bibitem{hin92a} \ref{M. Hindmarsh} {\prl} 68 1263 1992 .
\bibitem{hin92b} M. Hindmarsh, Cambridge/ITP preprint
DAMTP-HEP-92-24/NSF-ITP-92-75 (1992).
\bibitem{pre92} J. Preskill, CalTech preprint CALT-68-1787 (1992).
\bibitem{gib+} G. Gibbons, M. E. Ortiz, F. Ruiz Ruiz, and T. M. Samols,
Cambridge preprint DAMTP-R-92/7 (1992).
\bibitem{abr} E. Abraham, Cambridge preprint DAMTP-R-92/12 (1992).
\bibitem{gal} \ref{K. Galicki} {\np} B271 402 1986 .
\bibitem{gur-tze} \ref{F. G\"ursey and H. C. Tze} {Ann. Phys.} 128 29 1980 .
\bibitem{bre-hik-zin} \ref{E. Br\'ezin, S. Hikami, and J. Zinn-Zustin}
{\np} B165 528 1980 .
\bibitem{vac92} \ref{T. Vachaspati} {\prl} 68 1977 1992 .
\bibitem{FibBun} C. Nash and S. Sen, ``Topology and
Geometry for Physicists'' (Academic Press, London, 1983);
N. Steenrod ``The Topology of Fibre Bundles'' (Princeton
University Press, Princeton, 1951).
\bibitem{egu-gil-han} \ref{T. Eguchi, P. B. Gilkey, and A. J. Hanson}
{\prep} 66 213 1980 .
\bibitem{gil} R. Gilmore, ``Lie Groups, Lie Algebras, and some of their
applications'' (Wiley, NY, 1974).
\bibitem{raj} R. Rajaraman, ``Solitons and Instantons'' (North Holland,
Amsterdam, 1982).
\bibitem{tho} \ref{G. 't Hooft} {\pr} D14 3432 1976 .
\bibitem{aff} \ref{I. Affleck} {\np} B191 429 1981 .
\bibitem{Sphal} \ref{N. Manton} {\pr} D28 2019 1983 ;
\ref{F. Klinkhamer and N. Manton} {\pr} D30 2212 1984 .

\end{thebibliography}
\end{document}